\documentclass[aps,prx,10pt,twocolumn,superscriptaddress,longbibliography,nobibnotes,floatfix]{revtex4-2}
\usepackage[table]{xcolor}
\usepackage{tabularray}
\usepackage{xcolor}

\usepackage{graphicx}
\usepackage{dcolumn}
\usepackage{bm}
\usepackage{lineno}
\usepackage{amsmath}

\usepackage{placeins}
\usepackage{xcolor}
\usepackage{multirow}
\usepackage{makecell}
\usepackage{amssymb}
\usepackage{booktabs, multirow, siunitx}
\usepackage[normalem]{ulem}
\usepackage[colorlinks=true, allcolors=blue]{hyperref}

\usepackage[shortlabels]{enumitem}
\newcommand{\ECNU}{Quantum Institute for Light and Atoms, School of Physics, East China Normal University, Shanghai 200062, China}
\newcommand{\HE}{School of Physics, Henan Normal University, Xinxiang, Henan 453007, China}
\newcommand{\HEP}{School of Mathematics and Statistics , Henan Normal University, Xinxiang, Henan 453007, China}
\newcommand{\CHONG}{Chongqing Institute of East China Normal University, Chongqing 401120, China}
\newcommand{\SHAN}{Collaborative Innovation Center of Extreme Optics, Shanxi University, Taiyuan 030006, China}

\begin{document}
\title{Fast Adiabatic Quantum Gates via Hyperfine Intermediate States}

\author{Jiayin Fan}
\affiliation{\ECNU}
\affiliation{\HE}

\author{Xingdong Zhao}
\email{phyzxd@gmail.com}
\affiliation{\HE}

\author{Manqi Zhang}
\affiliation{\ECNU}
\affiliation{\HE}

\author{Fangfang Xie}
\affiliation{\HEP}

\author{Jing Qian}
\email{jqian1982@gmail.com}
\affiliation{\ECNU}
\affiliation{\CHONG}
\affiliation{\SHAN}

\date{\today}
\begin{abstract}
The appeal of adiabatic quantum computing lies in its intrinsic robustness against various technical imperfections, making it attractive for many quantum information applications. However, it faces a fundamental challenge: accelerating the adiabatic operations while preserving adiabaticity within the qubit coherence time. In this article, we propose an electromagnetically induced transparency-based adiabatic CNOT gate protocol which harnesses atomic hyperfine intermediate states (HISs) to speed up the adiabatic evolution. The HISs, naturally-existed in two-photon transitions, often need to be suppressed due to their significant decay errors. In contrast, this paper introduces a novel method that utilizes appropriately chosen HISs not only to enhance the adiabaticity in STAY pathway but also to accelerate the population transfer in TRANSFER pathway. Through pulse optimization, we achieve adiabatic gate fidelities exceeding 0.9991 within 0.3903 $\mu$s in realistic Cs atomic setups. To demonstrate the generality of protocol we further assess the impact of decays from multiple HIS and extend our model to arbitrary number of states, providing a practical route toward fast and robust adiabatic quantum gates in Rydberg-atom platforms.
\end{abstract}

\maketitle

\maketitle

\section{introduction}

Quantum gates based on adiabatic evolution have attracted intensive interests thanks to their inherent robustness against experimental imperfections \cite{Bacon2009,Beterov:2016,Frees2019,Setiawan2023,Xue2024}, and have been implemented in a variety of platforms \cite{Duan2001,Barends2016,Liang2016,Liu2024}. Unlike the dynamical gates that rely on a precise modulation of pulse amplitudes and phases, adiabatic gate protocols often require the follow of instantaneous eigenstates of a slowly varying Hamiltonian \cite{Bergmann1998}, thus rendering them insensitive to moderate fluctuations in laser intensity and pulse duration \cite{Goerz2014,Saffman2020,Mitra2020,Zheng2022,Turyansky2024}. 
This resilience is particularly valuable in Rydberg-atom platforms, where long-range interactions and strong dipolar moments enable the creation of 
high-fidelity entanglement, known as a central property of quantum mechanical systems \cite{Moller2008,Mitra2023,Yuan2026Beating}. 
However, the robustness of adiabatic quantum gates comes at the price of an extremely long operation time, as the speed of adiabatic evolution must remain slow compared to the energy gap between dark and bright eigenstates. In typical Rydberg excitation schemes, this can lead to microsecond-scale gate durations \cite{Rao:2014,Keating2015,Petrosyan2017,Wu:2017,Khazali2020}. Therefore, acceleration of adiabatic gate protocols has become an intriguing research focus \cite{Yu2019,Li2021,Rus2026,Turyansky2026}.


So far, a widely used acceleration technique is called shortcuts to adiabaticity (STA) \cite{Chen2010} which aims to engineer pulses that completely cancel the nonadiabatic errors \cite{GuerryOdelin2019}. While exact STA is highly effective its implementation requires the knowledge of all eigenstates and their derivatives making it practical only for relatively simple quantum operations \cite{Berry2009,Ibanez2012Multiple,delCampo2013,Claeys2019}. In many realistic systems the eigenstates are generally unknown, and the unwanted couplings to higher energy levels cannot be ignored even if approximate STA solutions are used, leading to non-negligible nonadiabatic errors \cite{Campbell2015,Sels2017,Hartmann2020,Barone2024}. For this reason a practical method that accelerates adiabatic quantum gates without exacerbating nonadiabatic errors, has remained challenging.

In parallel, the intermediate-state scattering from hyperfine intermediate states (HISs) is treated as a severe error source for quantum gate operations due to their short lifetimes \cite{Cong2022,Gerasimov2022,Evered2023}. A common way to suppress this scattering error involves increasing the intermediate-state detuning which substantially reduces the two-photon Rabi frequency (typically far below $ 2\pi\times 10$ MHz \cite{Fu2022}) resulting in very long gate operation times \cite{Ozeri2007,Zhang2012}. Recent achievement using high one-photon Rabi frequencies reaches a peak of $2\pi\times 17$ MHz albeit demanding extremely advanced techniques \cite{Evered2026}. Thereby, there exists a fundamental trade-off between the intermediate scattering error suppression and fast quantum gates. A natural question arises: can this challenging problem be resolved by employing multilevel quantum control techniques or advanced optimization methods?

In this work we present a new acceleration method for implementing a native electromagnetically induced transparency (EIT)-based CNOT gate between two alkali atomic qubits. This protocol having been demonstrated experimentally \cite{McDonnell2022EIT}, offers a scalable route to multiqubit CNOT$^k$ gates \cite{Farouk2023CNOTk}; yet it typically operates on a few microseconds timescale. Because, limited by the laser power, the need for coherent state transfer under EIT resonance results in longer pulse durations which essentially restricts the attainable gate fidelity even below 0.9. Here, we propose an acceleration mechanism by using a series of HISs, naturally existed in realistic two-photon excitation systems being a major source of decoherence \cite{Tate2018,Pelegri2022,Duspayev2025} . We identify that, when HISs couple to the ground and Rydberg states with appropriate strengths for the target atom, the dark-state evolution in STAY pathway can be enhanced due to a larger energy gap. Besides, the Raman laser driving the two-photon transitions of two ground manifolds in TRANSFER pathway is also shortened at the same time. This dual improvement offers a win-win enhancement for this gate protocol.

Crucially, our approach counterintuitively utilizes HISs that actively participate in the state dynamics, thereby facilitating the performance in 
both the STAY and TRANSFER pathways. We generalize the early work by M\"uller {\it et.at.} \cite{Muller2009} and focus on 
experimentally practical settings with HISs to realize fast adiabatic CNOT gates.
The pulse optimization method \cite{Theis2016,Pagano2022,Polat2026} is applied merely for the target atom leaving the control-atom simply driven by $\pi$-pulses. Even under the presence of all HISs and Rydberg-state decay errors our enhanced CNOT gate estimated for Cs atoms achieves gate fidelities exceeding 0.9991 with a gate time of 0.3903 $\mu$s.
We perform sufficient optimization runs to demonstrate that the advantages of our gate protocol can be robustly preserved when extended to systems with an arbitrary number $N$ of appropriate HISs and to other alkali atomic species. Our results offer an alternative method for achieving fast and robust adiabatic quantum gates in experimentally-feasible neutral‑atom platforms \cite{Li2022,Beterov2026,Rej2026}.

\section{Enhanced EIT-based adiabatic gates} \label{enh}


\begin{figure}
    \centering
    \includegraphics[width=7.3cm]{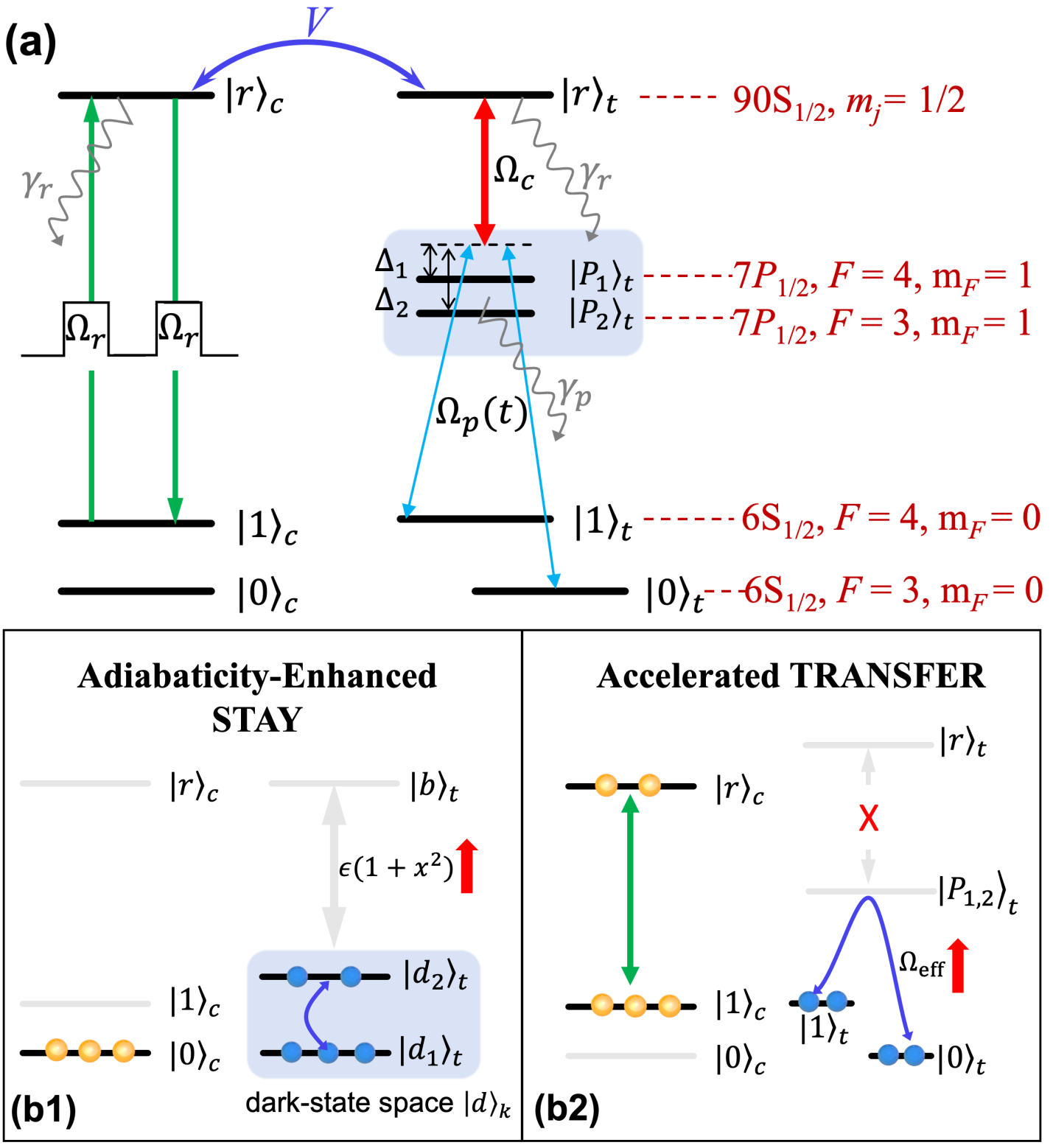}
    \caption{Illustration of EIT-based fast adiabatic CNOT gates with HISs. (a) shows the atomic energy levels in {\it e.g.}, Cs atoms. The control qubit atom is modeled as an effective two-photon system between $|1\rangle_c$ and $|r\rangle_c$ with $|0\rangle_c$ un-excited, which is operated by two separate square-$\pi$ pulses (labeled $\Omega_r$) serving as the control knob. The target qubit forming of an inverted-Y five-level configuration exhibits the excitation to $|r\rangle_t$ from two ground states $|0\rangle_t$ and $|1\rangle_t$ mediated by HISs $|P_1\rangle $ and $ |P_2\rangle$, detuned by $\Delta_1$ and $\Delta_2$, respectively. $\gamma_{p,r}$ denote the spontaneous decay rates. See Table \ref{tab:k_factors} for more detailed parameters. (b1,b2) illustrate the STAY and TRANSFER pathways, corresponding to the control atom in $|0\rangle_c$ and $|1\rangle_c$ initially, both of which are enhanced (adiabaticity-enhanced or accelerated) due to the presence of HISs, promising for constructing fast and adiabatic quantum gates.}
    \label{multimode}
\end{figure}

Our approach makes use of the original scheme \cite{Muller2009} as in shown Fig.\ref{multimode}a, while extending it with HISs to model the current protocol. 
The control qubit atom is implemented individually as a standard three-level structure consisting of the
ground states $|0\rangle_c, |1\rangle_c$ and a Rydberg state $|r\rangle_c$, driven from $|1\rangle_c \to |r\rangle_c$ by a two-photon square-$\pi$ pulse
with Rabi frequency $\Omega_r$, whereas atoms in $|0\rangle_c$ remain
unexcited. For the target qubit atom, we model it
as a native five-level system in which a pair of $\sigma^+$-polarized
Raman lasers with Rabi frequencies $\Omega_p(t)$ drives a two-photon resonance from $|0\rangle_t \to |1\rangle_t$, with a detuning $\Delta_1$ relative to the HIS $|P_1\rangle$. A strong coupling
laser with detuning $-\Delta_1$ couples $|P_1\rangle \to |r\rangle_t$ with Rabi
frequency $\Omega_c$. 
More specifically, given that 
we consider the excitation via $7P_{1/2}$ state in Cs atoms, this choice will practically necessitate the inclusion of another hyperfine state $|P_2\rangle$ due to the allowed dipole transitions (see the energy-level labelings on the right).
Because
the
linewidth of excitation lasers is often larger
than the intermediate-state splittings leading to the simultaneous coupling of multiple hyperfine manifolds with single Raman pulse \cite{Naber2016}. Remarkably, the participation of $|P_2\rangle$ arises different coupling strengths
$\{k_0\Omega_p$, $k_1\Omega_p$, $k_r\Omega_c\}$ corresponding to the states $|0\rangle_t$, $|1\rangle_t$,
$|r\rangle_t$, detuned by $\Delta_2$, where the factor $k$ reflects the relative 
strength of dipole matrix
elements and $\Delta_{21}=|\Delta_2-\Delta_1|$ represents the hyperfine splitting energy. 

Thereby, the full Hamiltonian describing this two-qubit gate system, can be
\begin{equation}
    \mathcal{H} = \mathcal{H}_{c} \otimes I  + I \otimes \mathcal{H}_{t} + V|rr\rangle \langle rr|,
\end{equation}
where $c,t$ label the control and target atoms, and 
\begin{eqnarray}
    \mathcal{H}_c &=& \frac{\Omega_r}{2}|1\rangle_{c}\langle r| + \text{H.c.}, \\
\mathcal{H}_t & =&  \sum_{\alpha=0,1}\frac{\Omega_{p}(t)}{2}|\alpha\rangle_t\langle P_1| + \frac{\Omega_c}{2} |P_1\rangle_t\langle r| + \text{H.c.}  \nonumber \\
&+&  \sum_{\alpha=0,1}\frac{k_{\alpha}\Omega_{p}(t)}{2}|\alpha\rangle_t\langle P_2|+ \frac{k_r\Omega_c}{2} |P_2\rangle_t\langle r| + \text{H.c.} \nonumber  \\
&-& \Delta_1 |P_1\rangle\langle P_1| - \Delta_2 |P_2\rangle\langle P_2|. \label{Ht}
\end{eqnarray}
Here, $V$ is the Rydberg blockade energy that scales approximately as $r^{-6}$ for van der Waals interaction \cite{Walker2008}.

{\it Adiabaticity-enhanced STAY pathway.-} We first analyze an enhanced dark state arising from the presence of hyperfine $|P_1\rangle,|P_2\rangle$ states. Remember that, if the control atom stays  unexcited in $|0\rangle_c$, the system reduces to the 
picture of single target atom. In the limit of $|\Delta_{1,2}| \gg \{\Omega_{p,c},k_{0,1}\Omega_p,k_r\Omega_c\}$ the hyperfine states $|P_1\rangle$ and $|P_2\rangle$ can be adiabatically eliminated. The target-atom Hamiltonian $\mathcal{H}_t$ is then described by an effective form in the $\{|0\rangle_t, |1\rangle_t, |r\rangle_t\}$ basis,
\begin{equation}
\mathcal{H}_{\text{eff}}^t = \frac{1}{4} \left( \frac{1}{\Delta_1} \mathbf{V}_1^T\mathbf{V}_1 + \frac{1}{\Delta_2} \mathbf{V}_2^T\mathbf{V}_2 \right), \label{Heff}
\end{equation}
with vectors defined as $\mathbf{V}_1 =\left( \Omega_p , \Omega_p , \Omega_c \right)$ and $\mathbf{V}_2 = \left( k_0\Omega_p, k_1\Omega_p, k_r\Omega_c \right)$. When 
\begin{equation}
    k\equiv k_0=k_1=k_r \label{kfac},
\end{equation}
known as the $k$-factor condition (see Sec. \ref{sec:k_factor}),
we can obtain a simpler form exactly same as the original one-intermediate-state case \cite{Muller2009}, in which the new basis vectors \{$|+\rangle_t = \frac{1}{\sqrt{2}}(|0\rangle_t+|1\rangle_t),|r\rangle\}$ are used,
\begin{equation}
\mathcal{H}_{\text{eff}}^t/\epsilon =  {x}^2 |+\rangle_{t}\langle+| + |r\rangle_{t}\langle r| + {x} \bigl( |+\rangle_{t}\langle r| + \text{H.c.} \bigr).  \label{efft}
\end{equation}
Note, another state $|-\rangle_t = \frac{1}{\sqrt{2}}(|0\rangle_t-|1\rangle_t)$ is decoupled and absolutely dark. Interestingly, $\mathcal{H}_{\text{eff}}^t$ offers an improved $k$-dependent energy scale $\epsilon = \frac{\Omega_c^2}{4} \left( \frac{1}{\Delta_1} + \frac{k^2}{\Delta_2} \right) > \frac{\Omega_c^2}{4\Delta_1}$ in which the dimensionless coefficient stays $
{x}(t) = \frac{\sqrt{2}\Omega_p(t)}{\Omega_c}$. For $x(t)\ll 1$, 
$\mathcal{H}_{\text{eff}}^t$ describes the EIT scenario that supports two zero-energy dark eigenstates
\begin{equation}
     |d_1\rangle_t = |-\rangle_t, \quad \quad   |d_2\rangle_t = \frac{1}{\sqrt{1+{x}^2} }\bigl( |+\rangle_t - {x} |r\rangle_t \bigr), \label{darkstate}
\end{equation}
as well as one bright eigenstate $|b\rangle_t = \frac{1}{\sqrt{1+{x}^2} }\bigl(  {x} |+\rangle_t +|r\rangle_t \bigr)$ with energy $\epsilon(1+x^2)$. Here, for the control atom initially in state $|0\rangle_c$, two $\pi$-pulses have no effect. The Raman adiabatic pulse $\Omega_p(t)$ together with $\Omega_c$ drives the system along the composite dark state $|d\rangle_k = \frac{1}{\sqrt{2}}(|d_1\rangle_t+|d_2\rangle_t)$ within the dark-state space (see Fig.\ref{multimode}(b1)) ensuring that the target atom returns to its initial state after the adiabatic pulse is applied \cite{Vitanov2017}. This is known as the STAY pathway
\begin{equation}
    |0\rangle_c|j\rangle_t\leftrightarrow |0\rangle_c|j\rangle_t  \quad  \quad \text{    with    } j\in (0,1).
\end{equation}

Remarkably, we find the adiabaticity of dark state can be enhanced due to the presence of the HISs because the energy gap $\sim \epsilon(1+x^2)$ between dark $|d\rangle_k$ and bright $|b\rangle_t$ states is counterintuitively increased as a result of a larger $\epsilon$. This yields an adiabaticity-enhanced STAY-pathway for CNOT gates.

{\it Accelerated TRANSFER pathway.-} When the control atom is in $|1\rangle_c$ the first two-photon $\pi$ pulse transfers population to state $|r\rangle_c$ inducing a Rydberg blockade shift $V$ on the target atom \cite{Lukin2001}. Throughout this work, we assume $V/2\pi=1$ GHz by operating two Cs atoms separated by about 3.56 $\mu$m \cite{Sibalic2017ARC}. The target-atom Hamiltonian is then modified as
\begin{equation}
    \tilde{\mathcal{H}}_t =  \mathcal{H}_t+V|r\rangle_t\langle r|. \label{Htp}
\end{equation}
Once $V\gg\frac{\Omega_c^2}{4}(\frac{1}{\Delta_1}+\frac{k^2}{\Delta_2})$ is satisfied the EIT condition is broken as described in Fig.\ref{multimode}(b2), and the target atom undergoes transfer within the ground-state manifolds enabling the transition of
\begin{equation}
  |1\rangle_c |0\rangle_t  \quad \leftrightarrow  \quad |1\rangle_c|1\rangle_t,
\end{equation}
via a smooth Raman pulse with area $\int_0^{\tau} dt\Omega_{\text{eff}}=\pi$ \cite{Xing2026}. Remarkably, with the help of HIS, the effective two-photon Rabi frequency also becomes $k$-dependent, given by
\begin{equation}
\Omega_{\text{eff}} = \frac{\Omega_p^{2}}{2}\left( \frac{1}{\Delta_1} + \frac{k^2}{\Delta_2} \right)> \frac{\Omega_p^2}{2\Delta_1},\label{eff}
\end{equation}
being explicitly enhanced which in turn leads to a shortened adiabatic pulse duration thereby accelerating the gate execution.

With these win-win improvements in both STAY and TRANSFER pathways the native hyperfine-state model (Fig.\ref{multimode}a) achieves an enhanced gate fidelity for the universal EIT-based CNOT gate scheme in theory. However, errors arising from multiple intermediate state scatterings may also increase, leading to additional gate imperfections in practice \cite{Pelegri2022}. In this work we employ the single-pulse optimal control along with increasing intermediate-state detuning to suppress this error.

\section{Implementation setup}

\begin{table*}
	\centering
	\caption{Summary of realistic atomic levels, dipole matrix elements, detunings and the corresponding $k_\alpha$ parameters (denoting the ratio of transition probabilities between $|\alpha\rangle$ and $|P_{1,2}\rangle$) for Cs and Rb atoms. The nuclear spin values are $I=7/2$ and $ 3/2$, respectively, and $\Delta_{21}$ represents the hyperfine splitting of detunings.}
    \label{tab:k_factors}
    \setlength{\tabcolsep}{10pt}
	\begin{tabular}{@{} l c c c c c c c
			  @{}}
		\toprule
        \multicolumn{8}{c}{\textbf{Cs} \quad ($I=7/2$)}\\
\midrule
\multicolumn{8}{c}{$|P_1\rangle = |7P_{1/2},F=4,m_F=1\rangle$, \text{   }$|P_2\rangle=|7P_{1/2},F=3,m_F=1\rangle$}\\
\midrule
		$|\alpha\rangle$ & Level & $|F,m_F\rangle$ & $|m_J\rangle$ &
$|\langle \alpha| e r |P_1\rangle|$ &$|\langle \alpha| e r |P_2\rangle|$  & $k_\alpha$&$\Delta_{21} /2\pi$(GHz) \\
		\midrule
		$|0\rangle_t$ & \multirow{2}{*}{$6S_{1/2}$} & $|3,0\rangle$ & / &0.088982  & 0.068925   &$0.775$ & \multirow{4}{*}{$0.060$} \\
        
		$|1\rangle_t$  & & $|4,0\rangle$ & / &0.088982  & 0.068925  &$0.775$  &    \\
         \multirow{2}{*}{$|r\rangle_t$} &\multirow{2}{*}{$90S_{1/2}$}&/ &$1/2$  & 0.003148 & 0.002438 & $0.775$ &\\
         & & / & $-1/2$  &  0.002438& 0.003148 & $1.291$ & \\        
\midrule
  \multicolumn{8}{c}{\textbf{Rb} \quad ($I=3/2$)}\\
  \midrule
\multicolumn{8}{c}{$|P_1\rangle =  |6P_{1/2},F=2,m_F=1\rangle$, \text{   }$|P_2\rangle =  |6P_{1/2},F=1,m_F=1\rangle$}\\
       \midrule
		$|\alpha\rangle_t$ & Level & $|F,m_F\rangle$ & $|m_J\rangle$ &
$|\langle \alpha| e r |P_1\rangle|$ &$|\langle \alpha| e r |P_2\rangle|$  & $k_\alpha$&$\Delta_{21} /2\pi$(GHz) \\
		\midrule
		$|0\rangle_t$ & \multirow{2}{*}{$5S_{1/2}$} & $|1,0\rangle$ & / &  0.117733 & 0.067973 &$0.577$ & \multirow{6}{*}{$0.042$} \\
        
		$|1\rangle_t$  & & $|2,0\rangle$ & / & 0.117733 & 0.067973 &$0.577$ &    \\
         \multirow{4}{*}{$|r\rangle_t$} &\multirow{4}{*}{$90D_{3/2}$}&/ &$3/2$ & 0.007446 & 0.004299 & $0.577$ & \\
         & & / & $1/2$ &  0.002482& 0.004299 & $1.732$ & \\
         & & / & $-1/2$ &  0.004299& 0.002482 & $0.577$ & \\
         & & / & $-3/2$ &  0.004299& 0.007446 & $1.732$ & \\

	\bottomrule
	\end{tabular}
\end{table*}

\subsection{$k$-factor Condition}
\label{sec:k_factor}

To demonstrate an enhanced EIT adiabatic gate in practice we first identify the $k$-factor condition. Numerical estimates for Cs and Rb atoms are summarized in Table \ref{tab:k_factors}, adopted from accessible Rydberg experimental platforms \cite{Levine2019,McDonnell2022EIT}.
Here, in {\it e.g.}, Cs atoms as plotted by Fig.\ref{multimode}a, due to the nuclear spin $I=\frac{7}{2}$ we ensure that the available HISs are $|F=3,m_F=1\rangle$ and $|F=4,m_F=1\rangle$ for $7P_{1/2}$ manifold. The choice of $m_F=1$ depends on the dipole selection rules for the $\sigma^+$-polarized Raman laser $\Omega_p(t)$, which couples the ground and intermediate manifolds. Then, using the distinct dipole matrix elements calculated by the Wigner-Eckart theorem \cite{Sobelman2012}, we find that $k_0 = k_1$ holds when both transitions are driven by the same $\sigma^+$ (or $\sigma^-$) polarization alongside with appropriate $m_F$ matching. Notably, it fails if a $\pi$-polarized pulse is used.
Furthermore, whether $k_c=k_{0,1}$ holds depends on the polarization of the coupling laser $\Omega_c$ as well as the $m_J$ value of the Rydberg state $|r\rangle_t$. 
For the $\pi$-polarized $\Omega_c$ as adopted here, Table \ref{tab:k_factors} shows that $k_c\approx k_{0,1}=0.775$ when $ |r\rangle_t=|90S_{1/2},m_J=1/2\rangle$. However, the choice of $m_J=-1/2$ will lead to $k_c\neq k_{0,1}$ in Cs atoms (see the Failure case in Sec.\ref{fail}). For Rb atoms if the selected Rydberg state is $|r\rangle_t = |90D_{3/2},m_J=3/2\rangle$ we also find the satisfactory condition $k_c\approx k_{0,1}=0.577$ which can be used to achieve fast adiabatic quantum gates (see Appendix \ref{apa}).

\subsection{Numerical Verification}

To verify the presence of an enhanced dark state that can facilitate adiabatic quantum gate operations, we apply a trivial cosine-shaped smooth adiabatic pulse \cite{Rej2026a}
 \begin{equation}
     \Omega_p(t)=\frac{\Omega_p^{\max}}{2}\left[1-\cos(\frac{2\pi t}{\tau})\right],
     \label{nonop}
 \end{equation}
with which the population dynamics solely for the target atom in the two enhanced pathways, can be solved. The pulse duration $\tau$ is automatically determined by the $\pi$-area restriction in the TRANSFER pathway.
Note that the target atom initially resides in $|0\rangle_t$; the dynamics of input $|1\rangle_t$ exactly mirrors this as $|1\rangle_t$ resonantly couples to intermediate states with the same coupling strength $\Omega_p(t)$. All simulations use a single-atom model governed by the Hamiltonian $\mathcal{H}_t$ (Eq.\ref{Ht}) for STAY and $\mathcal{H}_t^\prime$ (Eq.\ref{Htp}) for TRANSFER, resolving from the Schr\"odinger equation including dissipative terms ($\propto\gamma_{p,r}$). For comparison we also show the ideal dark state evolution following $|d\rangle_k$ in STAY which serves to validate the correctness of our theory and to demonstrate the excellent adiabaticity achieved.

\begin{figure}
\centering
\includegraphics[width=0.47\textwidth]{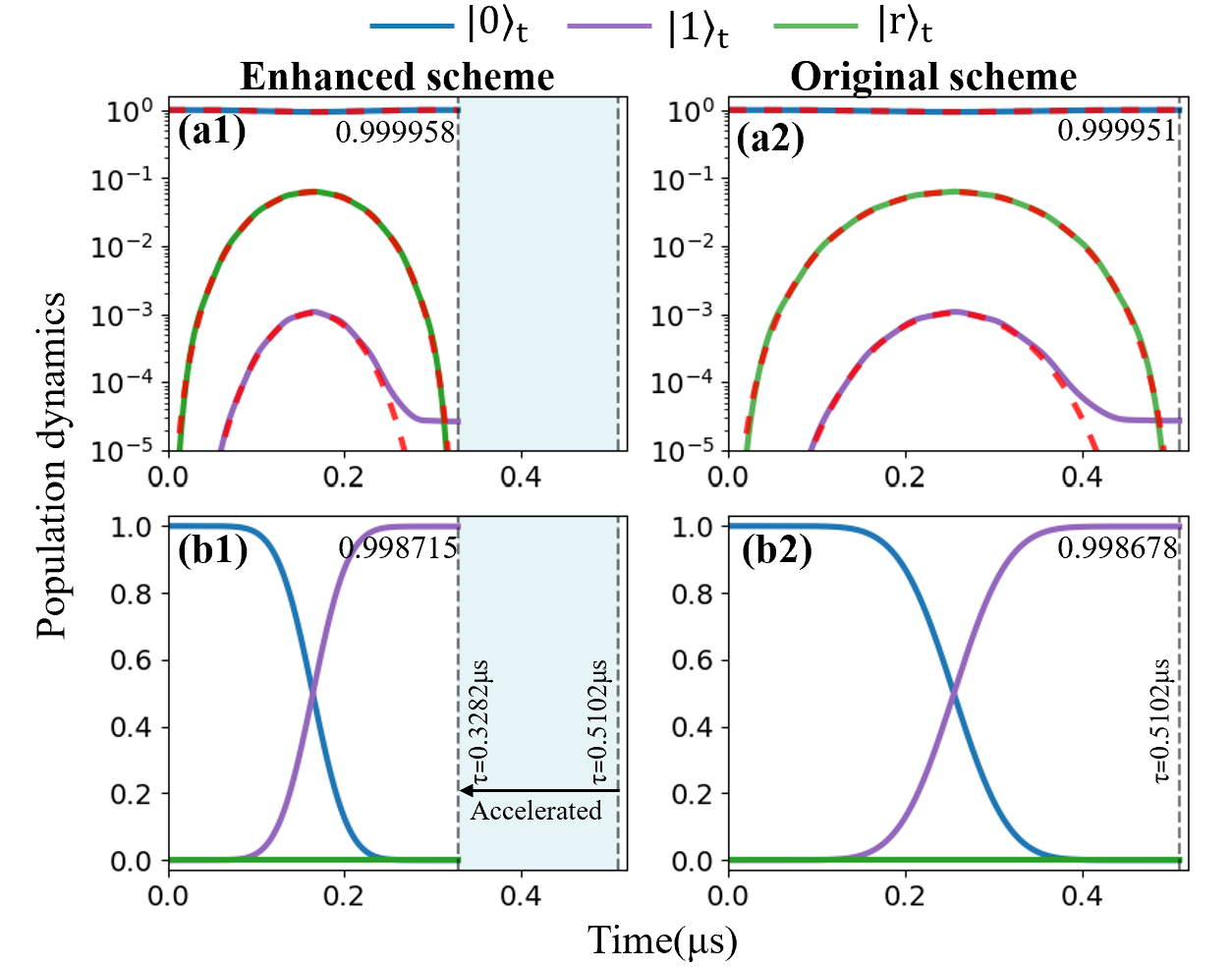}
\caption{Comparison of the target-atom population dynamics in the protocols with double (left, enhanced scheme) and single (right, original scheme) HISs. Panels (a1,a2) correspond to the STAY pathway and panels (b1,b2) show the TRANSFER pathway in which the final-target-state population presents the pathway fidelity. Numerical results, based on Schr\"odinger equation (with decays $\gamma_{p,r}$) and  analytical dark $|d\rangle_k$ state results (without decays), are given by solid and dashed curves respectively. In all plots the parameters are adopted from Cs atoms, which are $\Omega_p^{\max}/2\pi = 80$ MHz, $\Omega_c/2\pi = 300$ MHz, $\Delta_1/2\pi = 1.2$ GHz, $\Delta_2/2\pi = 1.26$ GHz, $k=0.775$,  $\gamma_r = 1.3$ kHz. Here we set a moderate decoherence rate $\gamma_p=6.5$ MHz for intermediate $P$ states, and this value will be tuned later (see Sec. \ref{int}) to study the role of multiple intermediate-state decay errors. }
\label{fig:adiabatic_following}
\end{figure}

As shown in Fig.\ref{fig:adiabatic_following}(a1,a2), benefiting from a second hyperfine intermediate state $|P_2\rangle$,
our enhanced scheme exhibits an evolution of the perfect dark state $|d\rangle_k$ (dashed curves), in agreement with the analytical results, just like in the original model with a single intermediate state. This indicates that the improved dark state maintains a high degree of adiabaticity, ensuring the STAY pathway for the target atom with a high fidelity $\sim 0.999958$. Note a similar value of $\sim$0.999951 is found for the original model. More remarkably,
although the adiabatic behavior of the STAY pathway remains nearly identical, our enhanced scheme significantly shortens the pulse duration $\tau$ which is essentially constrained by the Raman $\pi$ pulse requirement in the TRANSFER pathway.

For the TRANSFER pathway, thanks to the auxiliary intermediate state $|P_2\rangle$, the effective two-photon Rabi frequency $\Omega_{\text{eff}}$ between two ground states is always increased (see Eq.\ref{eff}) by a factor of $(1+k^2\Delta_1/\Delta_2) \approx 1.57>1$. This enhancement leads to a substantial reduction in the pulse duration $\tau$, as the $\pi$-pulse area condition $\int_0^{\tau} dt\Omega_{\text{eff}}=\pi$ must be satisfied. A direct comparison between Fig.\ref{fig:adiabatic_following}(b1) and (b2) explicitly demonstrates the advantage of our scheme
which achieves a much shorter pulse duration $\tau\approx 0.3282\mu$s compared to the original duration ($\tau\approx 0.5102\mu$s) while preserving a high transfer fidelity level $\sim 0.998715$ ($\sim0.998678$ originally). This result reflects a $35.7\%$ speedup for the adiabatic quantum gate, consistent with the analysis of the enhanced effective Rabi frequency  $\Omega_{\text{eff}}$. From an experimental standpoint, rather than requiring a longer duration, this enhanced adiabatic gate guided by naturally-existed HISs \cite{Rajasree2020,Li2023},
favors the shorter pulse and higher adiabaticity, promising for more reliable improvements under realistic experimental conditions.

\subsection{Pulse Optimization and Gate Characterization}
\label{sec:optimization}

Single-pulse optimization begins with the choice of a more tunable and smooth pulse waveform in the adiabatic regime that ensures the symmetric operation with respect to $t=\tau/2$ \cite{Sun2020Controlled},
\begin{equation}
\Omega_p(t) = \sum_{\nu=1}^{4} \beta_\nu \left[ b_{\nu,8}\left(\frac{t}{\tau}\right) + b_{8-\nu,8}\left(\frac{t}{\tau}\right) \right],
\end{equation}
where $b_{v,8}$ is the 8-order Bernstein basis polynomial and $\beta_{1\sim4}$ are the optimization coefficients. $\tau$ also labels the Raman pulse duration. The overall gate time is given by $T=\tau + 2t_\pi$ with $t_\pi=\pi/\Omega_r$ being the duration of the square $\pi$-pulse applied solely to the control atom. We use the genetic algorithm (GA) \cite{Katoch2020Review} for analysis and optimization of gate performance as in our recent work \cite{Li2022Optimal}. This algorithm is designed for ensuring a global search within 
the specified range of all tunable parameters. For a fair comparison with the original non-optimal cosine pulse (Eq.\ref{nonop}) we set a wide range for every $\beta$ coefficient with $\beta_{v}\in[0,\Omega_p^{\max}]$ and additionally constrain the peak value such that $\bar{\Omega}_p^{\max}=\max[\Omega_p(t)]\leq \Omega_p^{\max}$, and meanwhile decide the moderate duration within a relatively broad range $\tau\in[0,1.0]$ $\mu$s. Note that the values of  $\bar{\Omega}_p^{\max}$  and $\tau$ essentially have a trade-off relation because of the $\pi$-pulse-area condition; however, this trade-off is overcome by the choice of cost function $\mathcal{J}$ which has taken account of all decays in optimization.

In implementing GA, our optimization objective is to maximize the average gate fidelity $\mathcal{F}_0(\gamma_p,\gamma_r,T)$ (see definition in Eq.\ref{ff}) in the presence of all spontaneous decays from hyperfine intermediate and Rydberg states, so 
\begin{equation}
  \mathcal{J} = \mathcal{F}_0(\gamma_p,\gamma_r,T),
\end{equation}
serves as the cost function. This strategy allows us to set the population size to be 60 and the maximal generation number to be 120, sufficiently balancing the computational overhead and the search diversity in a five-dimensional parameter space $\{\beta_{1\sim4},\tau\}$ \cite{Mitchell1998}. Here 120 generations is selected based on the convergence test where we observe that the fitness value $\mathcal{J}$ typically stabilizes within 100 generations, and the additional 20 generations contribute a safe margin to ensure full convergence. Initially, a population of 60 candidate solutions is randomly generated in which each candidate consists of a set of five parameters. During every generation the cost function $\mathcal{J}$ is evaluated by computing the final-time outcome through full numerical simulation of the gate operation. At last, the best-performing candidates are selected for reproduction, and crossover and mutation operations are applied to generate the next generation. This process is repeated for up to 120 generations, and the best candidate from all generations is taken as the optimized solution.

To characterize the practical gate performance it is convenient to define a gate fidelity based on system's evolution. To account for the spontaneous decays from hyperfine intermediate and Rydberg states we describe the full system using non-Hermitian Hamiltonians
\begin{equation}
    \mathcal{H}^\prime= \mathcal{H}_{c}^\prime \otimes I  + I \otimes \mathcal{H}_{t}^\prime + V|rr\rangle \langle rr|,
\end{equation}
with 
\begin{eqnarray}
    \mathcal{H}_c^\prime &=& \mathcal{H}_c-  \frac{i\gamma_r}{2}|r\rangle_c\langle r|, \nonumber\\
    \mathcal{H}_t^\prime &=& \mathcal{H}_t - \frac{i\gamma_r}{2}|r\rangle_t\langle r| - \frac{i\gamma_p}{2}(|P_1\rangle_t\langle P_1|+|P_2\rangle_t\langle P_2|),\nonumber
\end{eqnarray}
having incorporated all dissipative terms ($\propto \gamma_{r,p}$) into the quantum dynamics. The time evolution of system is in turn governed by the Schr\"odinger equation
\begin{equation}
    i\frac{d|\Phi(t)\rangle}{dt} = \mathcal{H}^\prime |\Phi(t)\rangle,
    \label{sch}
\end{equation} 
 with $|\Phi(t)\rangle$ the full state vectors of two atoms. This approach provides a lower bound on the estimated fidelity, as the non-Hermitian terms cause population loss from the computational basis while neglecting the small probability of atoms returning to the logical qubit levels. 

Before defining the average gate fidelity $\mathcal{F}_0$ \cite{Shi2022} it is  useful to introduce two intermediate quantities that characterize the respective performance of STAY and TRANSFER pathways: $\mathcal{F}_s$ and $\mathcal{F}_t$. For the STAY fidelity $\mathcal{F}_s$, the control atom is initialized in $|0\rangle_c$ (the control \(\pi\)-pulse has no effect), and the target atom is prepared in \(|0\rangle_t\). After the time evolution, the probability that the target atom remains in \(|0\rangle_t\) is denoted as \(P_{0 \rightarrow 0}\). Due to the symmetry, $P_{1\to1} = P_{0\to0}$. The final state $|\Phi(T)\rangle$ is obtained by the Schr\"odinger equation (\ref{sch}) over the entire pulse sequence. The probability amplitude of the basis state \(|0\rangle_c \otimes |0\rangle_t\) is given by:
$c_{00} = \left( \langle 0|_c \otimes \langle 0|_t \right) |\Phi(T)\rangle$, therefore, the STAY fidelity is simply $\mathcal{F}_s = |c_{00}|^2 = P_{0\to0}$. Similarly, the TRANSFER fidelity is $\mathcal{F}_t = |c_{10}|^2 = P_{0\to 1}$ with $c_{10} = \left( \langle 1|_c \otimes \langle 0|_t \right) |\Phi(T)\rangle$ the probability amplitude according to the initial basis state \(|1\rangle_c \otimes |0\rangle_t\). Finally the average gate fidelity capturing how accurately the gate transforms arbitrary input state into the desired output state, defined as
 \begin{equation}
\mathcal{F}_{0} = \frac{1}{2} \left( \mathcal{F}_s + \mathcal{F}_t \right), \label{ff}
\end{equation}
assuming that the performance is symmetric with respect to the target atom's initial state $|0\rangle_t$ or $|1\rangle_t$. This relationship emphasizes that both pathways contributing equally, must be optimized simultaneously to achieve the high-fidelity gate operation.

\begin{table*}
	\centering
	\caption{Gate performance between original single- and enhanced double-intermediate state schemes using different Raman pulses: non-optimized Cosine shaped (labeled C) and optimized Bernstein shaped (labeled B). We use $\Omega_r/2\pi=10$ MHz giving $t_\pi = 0.05\mu$s and the overall gate time is $T(\mu s)=\tau+0.1$. Other parameters are identical to those used in Fig.\ref{fig:adiabatic_following}. The average gate fidelity is calculated by $\mathcal{F}_0=(\mathcal{F}_s + \mathcal{F}_t)/2$ where $\mathcal{F}_{s,t}$ denote the pathway element fidelities. Note that only $\mathcal{F}_t$ (not $\mathcal{F}_s$) contains the Rydberg decay error ($\sim \gamma_r\tau$) arising from the control atom, therefore $\mathcal{F}_t\ll\mathcal{F}_s$. }
    \label{tab:performance_comparison}
	\begin{tabular}{@{} l c c c c c c
			S[table-format=1.5]
            c
			S[table-format=1.5] 
			S[table-format=1.5]
            S[table-format=1.5] 
			  @{}}
		\toprule
		 Case & Shape & \text{ }Scheme\text{ } & \text{ }$\bar{\Omega}_p^{\max}/2\pi$(MHz)\text{ } & \text{ }\text{ }$\frac{(\beta_1,\beta_2,\beta_3,\beta_4)}{2\pi}$ (MHz) \text{ } \text{ }&$\gamma_p$(MHz)\text{ }  &\text{ }$\tau$ (\(\mu\)s)\text{ }\text{ } & \text{ }\text{ }$\mathcal{F}_{s}$\text{ }\text{ } & \text{ }\text{ }$\mathcal{F}_{t}$ \text{ }\text{ }& \text{ }\text{ }$\mathcal{F}_{0}$\text{ }\text{ }   \\
		\midrule
		\multirow{4}{*}{I}
		&\cellcolor{blue!10} &\cellcolor{blue!10}enhanced &\cellcolor{blue!10}
        &\cellcolor{blue!10}
        & \cellcolor{blue!10} &\cellcolor{blue!10}0.3282 &\cellcolor{blue!10} 0.99996 &\cellcolor{blue!10} 0.99822 &\cellcolor{blue!10} 0.99909  \\
		&\cellcolor{blue!10}\multirow{-2}{*}{C} 
        &\cellcolor{blue!10}original &\cellcolor{blue!10}\multirow{-2}{*}{80.00} 
        &\cellcolor{blue!10}\multirow{-2}{*}{/}
        &\cellcolor{blue!10}\multirow{-2}{*}{6.50} &\cellcolor{blue!10}0.5102 &\cellcolor{blue!10}0.99995 &\cellcolor{blue!10}0.99795 &\cellcolor{blue!10} 0.99895 \\
		&\cellcolor{gray!15} &\cellcolor{gray!15}enhanced &\cellcolor{gray!15}77.20  &\cellcolor{gray!15}( 2.35,73.51,75.54,51.06) &\cellcolor{gray!15} &\cellcolor{gray!15}0.2903 &\cellcolor{gray!15} 0.99995 &\cellcolor{gray!15} 0.99827 &\cellcolor{gray!15} 0.99911  \\
		&\cellcolor{gray!15}\multirow{-2}{*}{B} 
        &\cellcolor{gray!15}original  &\cellcolor{gray!15}78.43 &\cellcolor{gray!15}(40.96,43.19,51.98,79.89)
        &\cellcolor{gray!15}\multirow{-2}{*}{6.50} &\cellcolor{gray!15}0.4310 &\cellcolor{gray!15} 0.99993 &\cellcolor{gray!15} 0.99805
        &\cellcolor{gray!15} 0.99899  \\
      
        \midrule
		\multirow{4}{*}{II}
		&\cellcolor{blue!10} &\cellcolor{blue!10}enhanced &\cellcolor{blue!10}
        &\cellcolor{blue!10}
        &\cellcolor{blue!10} &\cellcolor{blue!10}0.1463 &\cellcolor{blue!10} 0.99904 &\cellcolor{blue!10} 0.99847 &\cellcolor{blue!10} 0.99875  \\
		&\cellcolor{blue!10}\multirow{-2}{*}{C} 
        &\cellcolor{blue!10}original &\cellcolor{blue!10}\multirow{-2}{*}{120.00}  
        &\cellcolor{blue!10}\multirow{-2}{*}{/}
        &\cellcolor{blue!10}\multirow{-2}{*}{6.50}
        &\cellcolor{blue!10}0.2272 &\cellcolor{blue!10} 0.99898 &\cellcolor{blue!10} 0.99832 &\cellcolor{blue!10}0.99865  \\
		&\cellcolor{gray!15} &\cellcolor{gray!15}enhanced &\cellcolor{gray!15}89.41  &\cellcolor{gray!15}(22.75,50.40,36.80,111.29) &\cellcolor{gray!15} &\cellcolor{gray!15}0.2433 &\cellcolor{gray!15} 0.99992 &\cellcolor{gray!15} 0.99834 &\cellcolor{gray!15} 0.99913  \\
		&\cellcolor{gray!15}\multirow{-2}{*}{B} 
        &\cellcolor{gray!15}original  &\cellcolor{gray!15}87.97 &\cellcolor{gray!15}(23.60,49.23,53.00,96.07 )&\cellcolor{gray!15}\multirow{-2}{*}{6.50} &\cellcolor{gray!15}0.3801 &\cellcolor{gray!15} 0.99991 &\cellcolor{gray!15} 0.99812 &\cellcolor{gray!15} 0.99901  \\
      
        \midrule
		\multirow{10}{*}{III}
		
		&\cellcolor{gray!15} &\cellcolor{gray!15}enhanced &\cellcolor{gray!15}79.08  &\cellcolor{gray!15}(17.59,67.59,44.45,80.00) &\cellcolor{gray!15}
        &\cellcolor{gray!15}0.2802 &\cellcolor{gray!15} 0.99994 &\cellcolor{gray!15} 0.99700 &\cellcolor{gray!15} 0.99847  \\
		&\cellcolor{gray!15}\multirow{-2}{*}{B} 
        &\cellcolor{gray!15}original  &\cellcolor{gray!15}75.68  &\cellcolor{gray!15}(25.86,63.15,74.48,50.59)
        &\cellcolor{gray!15}\multirow{-2}{*}{13.00}  
        &\cellcolor{gray!15}0.4340 &\cellcolor{gray!15} 0.99993 &\cellcolor{gray!15} 0.99673 &\cellcolor{gray!15} 0.99833  \\
		&\cellcolor{gray!15} &\cellcolor{gray!15}enhanced &\cellcolor{gray!15}76.61  &\cellcolor{gray!15}(11.32,79.99,56.09,61.94) &\cellcolor{gray!15}
        &\cellcolor{gray!15}0.2823 &\cellcolor{gray!15} 0.99993 &\cellcolor{gray!15} 0.99572 &\cellcolor{gray!15} 0.99783  \\
		&\cellcolor{gray!15}\multirow{-2}{*}{B} 
        &\cellcolor{gray!15}original  &\cellcolor{gray!15}77.43 &\cellcolor{gray!15}(16.26,78.97,36.31,79.10)
        &\cellcolor{gray!15}\multirow{-2}{*}{19.50}
        &\cellcolor{gray!15}0.4372 &\cellcolor{gray!15} 0.99992 &\cellcolor{gray!15} 0.99541 &\cellcolor{gray!15} 0.99767  \\
		&\cellcolor{gray!15} &\cellcolor{gray!15}enhanced &\cellcolor{gray!15}78.07  &\cellcolor{gray!15}( 6.51,72.27,64.90,61.20) &\cellcolor{gray!15}
        &\cellcolor{gray!15}0.2858 &\cellcolor{gray!15} 0.99993 &\cellcolor{gray!15} 0.99444 &\cellcolor{gray!15} 0.99719  \\
		&\cellcolor{gray!15}\multirow{-2}{*}{B} 
        &\cellcolor{gray!15}original  &\cellcolor{gray!15}79.12 &\cellcolor{gray!15}(17.57,62.29,56.68,72.43)
        &\cellcolor{gray!15}\multirow{-2}{*}{26.00} 
        &\cellcolor{gray!15}0.4362 &\cellcolor{gray!15} 0.99992 &\cellcolor{gray!15} 0.99410 &\cellcolor{gray!15}0.99701   \\
		&\cellcolor{gray!15} &\cellcolor{gray!15}enhanced &\cellcolor{gray!15}79.50  &\cellcolor{gray!15}( 5.28,64.64,64.47,67.33) &\cellcolor{gray!15}
        &\cellcolor{gray!15}0.2884
        &\cellcolor{gray!15}0.99993 &\cellcolor{gray!15} 0.99317 &\cellcolor{gray!15} 0.99655  \\
		&\cellcolor{gray!15}\multirow{-2}{*}{B} 
        &\cellcolor{gray!15}original  &\cellcolor{gray!15}78.94 &\cellcolor{gray!15}(16.42,68.67,43.83,79.94)
        &\cellcolor{gray!15}\multirow{-2}{*}{32.50} 
        &\cellcolor{gray!15}0.4374  &\cellcolor{gray!15}0.99991 &\cellcolor{gray!15}0.99278  &\cellcolor{gray!15}0.99634   \\
		&\cellcolor{gray!15} &\cellcolor{gray!15}enhanced &\cellcolor{gray!15}77.55  &\cellcolor{gray!15}(13.50,71.17,61.91,62.28) &\cellcolor{gray!15}
        &\cellcolor{gray!15}0.2819 &\cellcolor{gray!15} 0.99991 &\cellcolor{gray!15} 0.99190 &\cellcolor{gray!15} 0.99591  \\
		&\cellcolor{gray!15}\multirow{-2}{*}{B} 
        &\cellcolor{gray!15}original  &\cellcolor{gray!15}77.63  &\cellcolor{gray!15}(19.19,59.44,80.00,51.99)
        &\cellcolor{gray!15}\multirow{-2}{*}{39.00} 
        &\cellcolor{gray!15}0.4348 &\cellcolor{gray!15} 0.99989 &\cellcolor{gray!15} 0.99148 &\cellcolor{gray!15} 0.99569  \\

		\bottomrule
	\end{tabular}
\end{table*}

\section{Gate performance}

\subsection{Comparing with One-Intermediate-State Gates}

To quantitatively demonstrate the advantages of enhanced gate scheme as well as the effectiveness of pulse optimization approach, we perform comparative simulations using cosine-type and Bernstein-type pulse shapes for both the enhanced (ours) and original schemes. Numerical results are summarized in Table \ref{tab:performance_comparison}. In Case I and II we initially set $\Omega_p^{\max}/2\pi=(80,120)$ MHz which leads to a fixed smooth cosine pulse (non-optimized) for two schemes and carry out the piecewise gates. As shown in the first row (marked in blue) of Cases I and II, we verify that the enhanced scheme exhibits significant advantages over the original one, which simultaneously favors a much shorter gate duration $\tau$ and a higher average fidelity $\mathcal{F}_0$. This improvement arises because both the STAY and TRANSFER pathways are enhanced by the auxiliary HISs in our protocol (see Sec. \ref{enh}). Furthermore, we observe that a large $\Omega_p^{\max}$ value may reduce the blockade constraint in STAY process leading to a relatively lower fidelity; therefore, $\Omega_p^{\max}/2\pi=80$ MHz is a preferable choice.

Further discussion focuses on the optimization case employing the Bernstein-type pulse, where the maximum value $\bar{\Omega}_p^{\max}$ is also constrained to $2\pi\times (80,120)$ MHz, accordingly. As shown in the second row of Cases I and II, although the allowed upper bound of $\bar{\Omega}_p^{\max}$ is significantly increased 
the optimized solution found by our algorithm remains moderate, striking a balance between the competing effects arising from fast gate duration and average gate fidelity \cite{Goerz2011}. In Case I where the upper bound is $\bar{\Omega}_p^{\max}/2\pi=80$ MHz, the optimization procedure favors a larger $\Omega_p(t)$ to suppress nonadiabatic leakage and intermediate-state scatterings \cite{Sun2014}. Moreover, accelerating the gate execution intuitively results in a lower $\mathcal{F}_0$ due to the breakdown of adiabaticity. Consequently, we observe that the optimal $\bar{\Omega}_p^{\max}$ obtained for a larger upper bound ($\bar{\Omega}_p^{\max}/2\pi=120$ MHz) remains much smaller than the bound itself, clearly confirming that the best optimized gate fidelity approaches $\mathcal{F}_0\approx 0.9991$. Remarkably, our enhanced gates, assisted by HISs, can outperform the original ones in both fidelity $\mathcal{F}_0$ ($0.99899\to 0.99911$) and gate duration $T$ ($0.5310\mu s\to 0.3903 \mu s$) truly enabling the realization of fast and adiabatic quantum gates.

\subsection{Larger Intermediate-State Decays} \label{int}

\begin{figure}
    \centering
\includegraphics[width=0.95\linewidth]{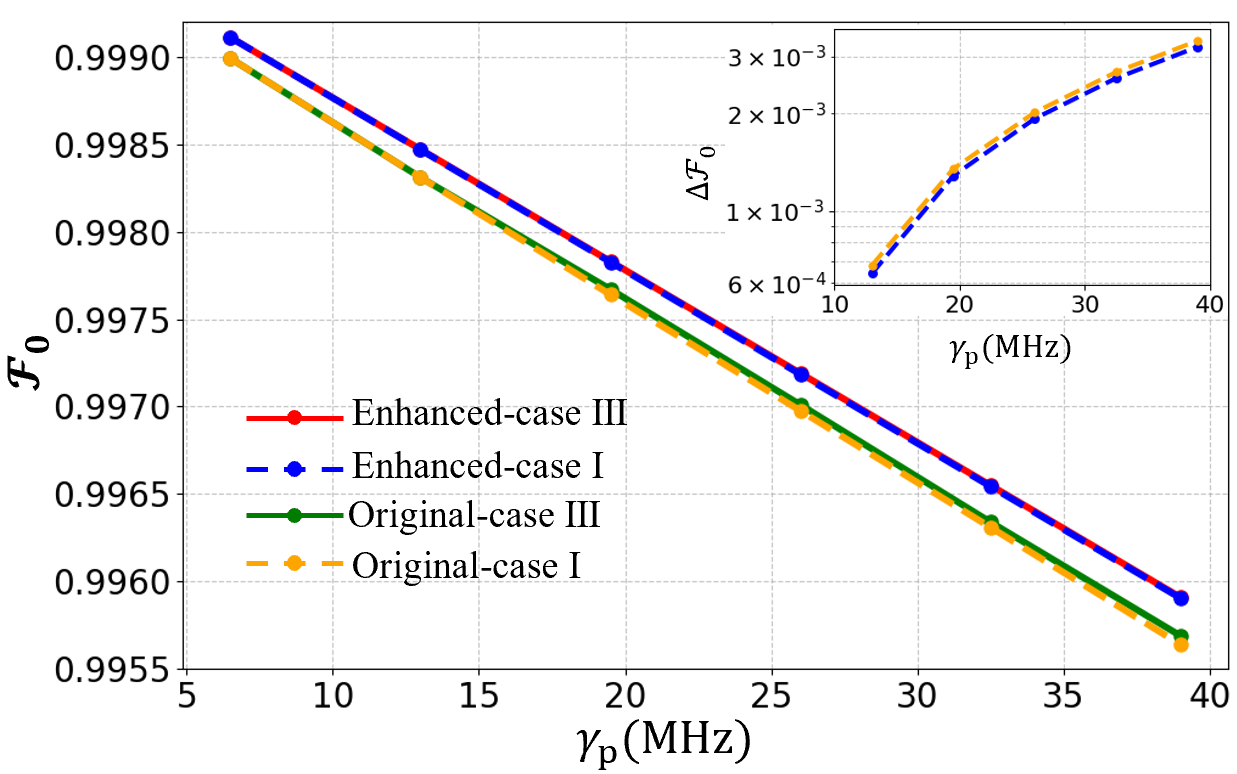}
    \caption{Average gate fidelity $\mathcal{F}_0$ as a function of the intermediate decay rate $\gamma_p$ based on the optimized Bernstein-shaped pulses. Solid curves are obtained via individual optimizations for every $\gamma_p$ value (Case III) while dashed curves are calculated by using the fixed optimized pulse at $\gamma_p=6.5$ MHz (Case I). Inset: Dependence of the derivation $\Delta\mathcal{F}_0=\mathcal{F}_0(\gamma_{p0})-\mathcal{F}_0(\gamma_p)$ on the decay rate $\gamma_p$ with $\gamma_{p0}=6.5$ MHz for Case I. }
    \label{fig:decay_robustness}
\end{figure}

Next, to test the robustness of gate performance against HIS scatterings we calculate the CNOT gate fidelity $\mathcal{F}_0$ across several different decay rates $\gamma_p$. Intuitively, decays from multiple HISs will introduce additional atomic loss, which reduces the advantages of our enhanced scheme. However,  
we show that employing single-pulse optimal control together with a shortened gate duration can effectively suppress the effect of decays from multi-intermediate states\cite{Pelegri2022,Evered2023}.

To quantify the influence of intermediate decays, in Fig.\ref{fig:decay_robustness} we intentionally increase $\gamma_p$ from its original value of $\gamma_p=6.5$ MHz up to 39.0 MHz, as indicated by the dashed curves (note that optimization depends on its original value). Although $\mathcal{F}_0$ generally decreases with increasing $\gamma_p$, our enhanced scheme reveals a smaller fidelity reduction. See the inset, the maximum derivation is $|\Delta\mathcal{F}_0|\approx 3.21\times 10^{-3}$ (blue-dashed), compared to $|\Delta\mathcal{F}_0|\approx 3.35\times 10^{-3}$ (orange-dashed) for the original scheme. This observation suggests that our enhanced scheme maintains stronger robustness and higher fidelities against multi-intermediate scattering errors, regardless of the decay rate magnitude.

Further results can be obtained by performing individual pulse optimizations for each decay rate, see Case III in Table \ref{tab:performance_comparison}. 
As expected, increasing 
$\gamma_p$ primarily reduces the transfer fidelity $\mathcal{F}_t$ which is affected by enhanced spontaneous scattering from the intermediate states.
In contrast, the STAY fidelity $\mathcal{F}_s$ remains close to unity ($\mathcal{F}_s \succeq 0.9999$) throughout the investigated parameter range, indicating that the dark-state adiabatic evolution is unaffected even under stronger intermediate-state dissipations. Remarkably, as shown in Fig.\ref{fig:decay_robustness} the results based on individual pulse optimizations almost overlap with those obtained from one-time original optimization. 
This indicates that the pulse shapes optimized in the weak-dissipation regime have been globally optimal.
Physically, pulse optimization primarily determines the coherent adiabatic dynamics, which is only weakly influenced by the intermediate dissipation strength. 
As the intermediate decay rate increases, its primary effect is to introduce additional scattering loss during the TRANSFER process without substantially modifying the optimal pulse shapes. Consequently, the optimized pulses exhibit strong robustness against variations in intermediate-state decay rates, highlighting the experimental feasibility of our enhanced scheme.

\section{Adiabaticity-enhanced failure }
\label{fail}

We next validate the severe necessity of the $k$-factor condition $k_0=k_1=k_r$ for constructing an enhanced adiabatic quantum gate. We employ the contrasting case where it is violated by replacing $|r\rangle_t$ with $m_J       =-1/2$, then one has $k_0=k_1=0.775$ and $k_r=1.291$ (Table \ref{tab:k_factors}, Cs atoms), providing a concrete example of the failure case.

When $k_0 = k_1 \neq k_r$, the coupling vector $\textbf{V}_1$ involved in constructing the effective Hamiltonian $\mathcal{H}_{\text{eff}}^t$ in Eq.(\ref{Heff}) is conserved, yet allowing $\textbf{V}_2$ to be expressed as
\begin{equation}
 \textbf{V}_2=(k\Omega_p,k\Omega_p,k_r\Omega_c).
\end{equation}

By introducing the energy scale $\epsilon^\prime=\frac{\Omega_c^2}{4}
\left(\frac1{\Delta_1}+
\frac{k_r^2}{\Delta_2}
\right)$ together with the dimensionless parameter $x=\frac{\sqrt2\,\Omega_p}{\Omega_c}$, the effective Hamiltonian in $\{|+\rangle_t,|r\rangle_t\}$ subspace
can be rewritten as,
\begin{equation}
\tilde{\mathcal{H}}_{\text{eff}}^{t}
=\epsilon^\prime\Bigl[a x^2 |+\rangle_t\langle+|
+|r\rangle_t\langle r|+b x\bigl(|+\rangle_t\langle r|
+\text{H.c.}\bigr)\Bigr],
\end{equation}
where
\begin{equation}
a=\frac{1+ \eta k^2}{1+\eta k_r^2},
\qquad
b=\frac{1+\eta k k_r}{1+\eta k_r^2},
\end{equation}
and $\eta = \frac{\Delta_1}{\Delta_2}$. Note that another state $|-\rangle_t = \frac{1}{\sqrt{2}}(|0\rangle_t-|1\rangle_t)$ is decoupled and absolutely dark, so $|d_1\rangle_t=|-\rangle_t$. When the $k$ factor condition $k_r=k$ is met, it yields $a=b=1$ resulting in $\tilde{\mathcal{H}}_{\text{eff}}^t = \mathcal{H}_{\text{eff}}^t $, which supports another zero-energy dark state $|d_2\rangle_t$ as given in Eq.~(\ref{darkstate}). By contrast, if $k_r\neq k$, $|d_2\rangle_t$ no longer exists, and we obtain a finite-energy (non-zero) dressed eigenstate
\begin{equation}
|\tilde d_2\rangle_t = \frac{1}{\sqrt{1+\tilde{x}^2}}
(|+\rangle_t-\tilde{x}|r\rangle_t),
\end{equation}
with $\tilde{x}=\frac{1}{2bx}\left[(a x^2-1)
+\sqrt{(a x^2-1)^2+4b^2x^2}\right]$ and the corresponding shifted eigenenergy expressed as
\begin{equation}
\mathcal{E}=\frac{\epsilon^\prime}{2}\left[(a x^2+1)-\sqrt{(a x^2-1)^2+4b^2x^2}
\right].
\end{equation}
Therefore, the STAY pathway can not ensure an exactly dark-state transfer and fails to return the population to the initial state, resulting in a reduced pathway fidelity $\mathcal{F}_s$.

\begin{figure}
    \centering
\includegraphics[width=0.95\linewidth]{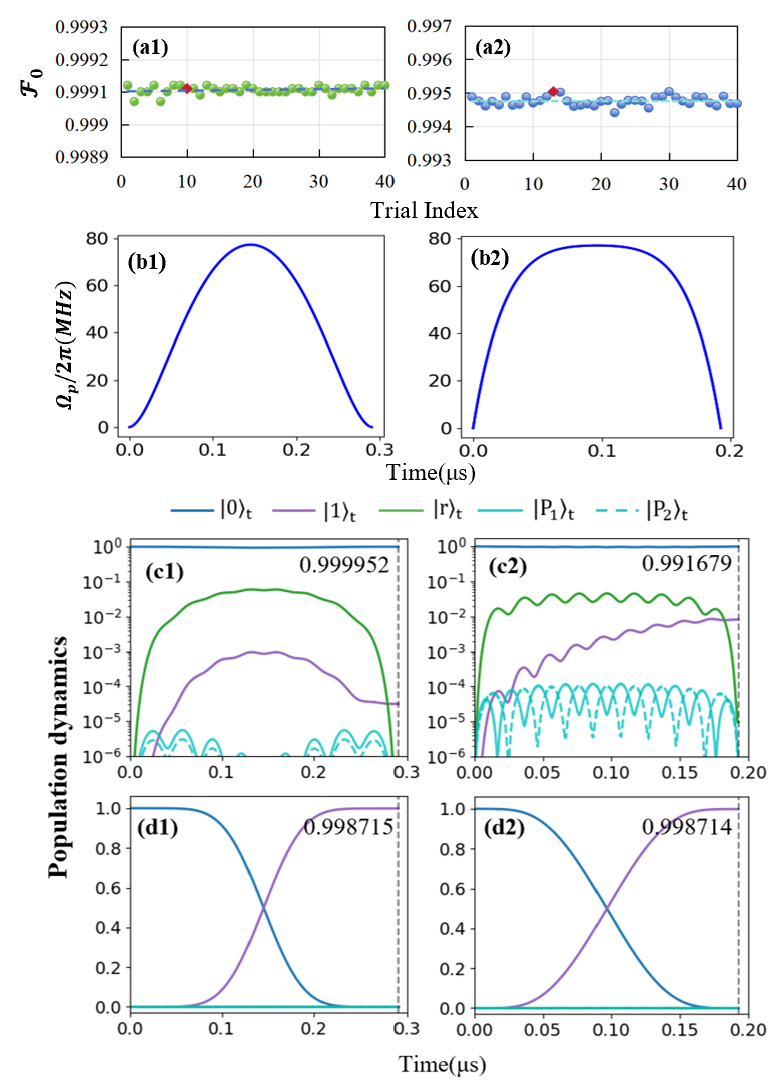}
    \caption{Gate performance for the adiabaticity-enhanced case ($k_r=k_{0,1}$) and the failure ($k_r\neq k_{0,1}$) case. (a1,a2) We show the average fidelity $\mathcal{F}_0$ over 40 independent optimization runs for two cases and respectively present (b1,b2) the optimized Raman pulse amplitude $\Omega_p(t)$ and (c1-2,d1-2) the two-pathway population dynamics for a specific solution, as highlighted by red diamonds in (a1) and (a2). Here, the specific-solution parameters are $\beta_{1\sim 4}/2\pi=(2.35,73.51,75.54,51.06)$ MHz which results in {$\tau=0.2903$} $\mu$s for the adiabaticity-enhanced case. While $\beta_{1\sim 4}/2\pi=(80.00,78.80,78.43,37.54)$ MHz, $\tau=0.1928$ $\mu$s for the failure case.}
    \label{Comoarision of optimizations} 
\end{figure}

 A quantitative discussion for the failure case is performed by comparing it with the adiabaticity-enhanced schemes. To account for the stochastic nature in optimization and ensure a fair comparison, we performed 40 independent optimization runs to evaluate the average gate fidelity $\mathcal{F}_0$ under two cases. The statistical results are displayed in Fig.\ref{Comoarision of optimizations}(a1) and (a2) which reveal that the best fidelity values achieved by the enhanced scheme remain above 0.999; however, they drop substantially to approximately $\sim 0.995$ level once the $k$ factor condition is unsatisfied. This fact confirms that the necessity of $k$ condition is not incidental.

 To further figure out the reason for the adiabaticity-enhanced failure we numerically compare the optimized pulse amplitude and two-pathway population dynamics in panels (b1-d1) and (b2-d2), using the marked optimized solutions. As expected, the adiabaticity-enhanced case remains high fidelities because of the presence of dark state that ensures the element fidelity $\mathcal{F}_s$ for STAY pathway to be as high as $\sim 0.99995$, via a strong suppression of population leakage to other intermediate excited or unwanted ground states. In contrast, owing to the breakdown of the dark state in the failure case, $\mathcal{F}_s$ for STAY pathway suffers from a significant reduction to be $\sim 0.9917$ arising from the effect of other intermediate unwanted states ({\it e.g.} $|1\rangle_t$ and $|P_{1,2}\rangle_t$). Notably, we observe that the optimized pulse has intelligently become a flat-top pattern (see (b2)) in an attempt to minimize the gate duration ($\tau$ is only 0.1928$\mu$s) while maintaining a high fidelity, just like the time-optimal strategy \cite{Jandura2022}; yet it is still unable to overcome the large spontaneous decay errors caused by the breakdown of dark state evolution in STAY. Therefore, although the TRANSFER pathway is hardly affected in the failure case, the overall gate fidelity is substantially decreased, falling far below 0.999.

 \section{Conclusion and outlook}

While adiabatic passage techniques have been demonstrated as a powerful tool for error-tolerant quantum gates, an intrinsic trade-off between gate speed and gate fidelity remains unsolved \cite{Song2024,Basilewitsch2024,Doultsinos2025}. Because adiabatic quantum gates often rely on an adiabatic state evolution to enhance their robustness against imperfections in control pulses, which in turn requires extremely long gate operation times making them potentially susceptible to dissipation \cite{Petrosyan2017,Mitra2023}. In this work, we have developed a protocol for realizing fast adiabatic quantum gates simply based on HISs, which naturally exist in realistic atomic two-photon systems \cite{McDonnell2022EIT}. With appropriate HISs that preserve the dark-state evolution we find that the adiabaticity for STAY pathway and the effective Rabi frequency for TRANSFER pathway can be simultaneously enhanced suggesting a promising route toward fast and high-fidelity adiabatic quantum gates.

As a demonstration of our protocol we analyze the implementation of an accelerated adiabatic CNOT gate in a realistic two-intermediate-level Rydberg Cs system, and focus on numerical optimization to obtain the best pulse shapes that mitigate both decay errors and nonadiabatic errors. Incorporating with realistic atomic levels and all spontaneous decay errors, we achieve a gate fidelity above 0.9991 within an overall gate duration of 0.3903 $\mu$s revealing an acceleration of approximately $\sim 26\%$. Moreover, our approach is very general and can be readily extended to systems with more HISs or to other atomic species such as Rb atoms. A rough estimate suggests that if more appropriate intermediate states exist they could lead to further speedup while preserving the gate fidelity at the same level.

To accelerate adiabatic quantum gates, 
our protocol counterintuitively uses HISs$-$whose effects are often avoided via large intermediate detunings in many two-photon transition settings \cite{Tretyakov2023}. Our results will hopefully prove to be an alternative and useful tool, beyond the STA method known as a promising solution to suppress nonadiabatic transitions by fast driving \cite{Baksic2016,Campbell2017,Alipour2020,Yin2022,CardenasLopez2023}, in various quantun-control related applications \cite{Khazali2020,Li2021,Wu2024}. Further research may explore how our theoretical ideas can be realized  experimentally even in the presence of technical noises \cite{Graham2019,Bluvstein2024,deLeseleuc2018} .

\begin{acknowledgments}

We acknowledge financial support from the National Natural Science Foundation of China under Grants Nos. 12174106, 11474094 and 11104076, the Natural Science Foundation of Chongqing under Grant No. CSTB2024NSCQ-MSX1117, the Shanghai Science and Technology Innovation Project under Grant No. 24LZ1400600, the Natural Science Foundation of Henan province under Grant No. 252300421995, Key Scientific Research Projects of Henan Province Higher Education Institutions Grant No. 26B140007 and the Science and Technology Commission of Shanghai Municipality under Grant No.18ZR1412800.

\end{acknowledgments}

\appendix

 \section{Extensive exploration results} \label{apa}

\begin{table*}
\centering
\caption{Optimized gate performance for extended systems with three and four HISs and Rb atoms ($N=2$). The last row presents the average value $\bar{\mathcal{F}}_0$ over 10 independent optimization runs to avoid the randomness in numerical optimization.}
\label{tab:multi_state}
\begin{tabular}{cccc}
\toprule
Parameters & $N=3$ & $N=4$ & Rb atom  \\
\midrule
$\frac{\bar{\Omega}_p^{\max}}{2\pi}$ (MHz) & 79.19 & 77.47  & 79.79 \\
$\frac{\beta_{1\sim 4}}{2\pi}$ (MHz)  & (17.41,32.14,73.53,71.14) & (18.20,47.80,51.80,79.03) & (23.84,78.45,72.58,53.74)\\
$\frac{\Delta_{1\sim N}}{2\pi}$ (GHz)  & (1.2,1.26,1.3) & (1.2,1.26,1.3,1.36) & (1.2,1.26)\\
$k_{2\sim N}$  & (0.775,1.2) & (0.775,1.2,0.6)  & (0.577)\\
all $\gamma_p$ (MHz)  & 6.5 & 6.5  & 7.7\\
\midrule
$\tau$($\mu$s)  & 0.1723 & 0.1565 & 0.2905 \\
$\mathcal{F}_{s}$  & 0.99997 & 0.99997  &  0.99992 \\
$\mathcal{F}_{t}$  & 0.99849 & 0.99853  &  0.99801 \\
$\mathcal{F}_{0}$  & 0.99923 & 0.99925 & 0.99897 \\
$\textcolor{black}{\bar{\mathcal{F}}_{0}}$  & {0.99920 $\pm$ 0.0000218 } & {0.99923$\pm$ 0.0000135} & {0.99896 $\pm$ 0.0000062 } \\
\bottomrule
\end{tabular}
\end{table*}

{\it More Hyperfine Intermediate States.-} In many application scenarios, the target qubit often couples to several hyperfine intermediate states. Here, we generalize it to a protocol with $N$ intermediate states, denoted as $|P_i\rangle$ ($i = 1, 2, \dots, N$). The corresponding couplings are characterized by Rabi frequencies $\Omega_{0i}$ and $\Omega_{1i}$ for the transitions $|0\rangle_t \leftrightarrow |P_i\rangle$ and $|1\rangle_t \leftrightarrow |P_i\rangle$, and $\Omega_{ci}$ for $|P_i\rangle \leftrightarrow |r\rangle_t$, respectively, with the intermediate detuning $\Delta_i$.

Within the rotating-wave approximation, we describe the general model by the target Hamiltonian
\begin{equation}
\begin{split}
\mathcal{H}_t = \frac{1}{2} \sum_{i=1}^{N} \Big( &\Omega_{0i} |0\rangle_t\langle P_i| + \Omega_{1i} |1\rangle_t\langle P_i| \\
&+ \Omega_{ci} |P_i\rangle_t\langle r| + \mathrm{H.c.}  \Big)
- \sum_{i=1}^{N} \Delta_i |P_i\rangle\langle P_i|.
\end{split}
\end{equation}
Assuming large detunings $|\Delta_i| \gg |\Omega_{0i}|, |\Omega_{1i}|, |\Omega_{ci}|$, these intermediate states can be adiabatically eliminated, yielding an effective Hamiltonian in the $\{|0\rangle_t,|1\rangle_t,|r\rangle_t\}$ basis :
\begin{equation}
\mathcal{H}_{\rm eff}^t = \sum_{i=1}^{N} \frac{1}{4\Delta_i} \mathbf{V}_i^T \mathbf{V}_i,
\end{equation}
where  $\mathbf{V}_1=
\begin{pmatrix}\Omega_p,\Omega_p,\Omega_c
\end{pmatrix}$ and $\mathbf{V}_i = \begin{pmatrix} \Omega_{0i},\Omega_{1i} ,\Omega_{ci} \end{pmatrix}$.
Note if all coupling vectors satisfy the $k$ factor condition with respect to $\mathbf{V}_1$, {\it i.e.}
\begin{equation}
\mathbf{V}_{i}=k_i\mathbf{V}_1,
\qquad i=2,\dots,N,
\end{equation}
one has $\Omega_{0i}=k_i\Omega_{01}$, $\Omega_{1i}=k_i\Omega_{11}$, $\Omega_{ci}=k_i\Omega_{c1}$. Furthermore, by imposing the symmetric Raman condition, we newly define $\Omega_{01}=\Omega_{11}=\Omega_p$, and $\Omega_{c1}=\Omega_c$ and obtain a new effective form
\begin{equation}
\mathcal{H}_{\rm eff}^t
=
\mu\,
\mathbf{V}_1^T\mathbf{V}_1,
\qquad
\mu=\frac{1}{4\Delta_1}+\frac{1}{4}
\sum_{i=2}^{N}
\frac{k_i^2}{\Delta_i}.
\end{equation}

We find that the system again supports two dark states in the $\{|+\rangle,|-\rangle\}$ basis, identical to those derived in Eq.~(\ref{darkstate}). Remarkably, any intermediate state, as long as meets the $k$-factor condition, can be involved that does not break the dark-state evolution. So we expect the win-win improvement in both STAY and TRANSFER pathways persists with more HISs.

In the following, we perform a numerical evaluation of gate fidelities in systems with $N=3$ and $4$ HISs while the $k$ factor condition is always satisfied. Note that the $k_3,k_4$ factors are arbitrarily chosen without corresponding to real energy levels.
The specific parameters and optimized results adopted are summarized in Table \ref{tab:multi_state}. All intermediate states have the same decay rate $\gamma_p$. Clearly, the adiabaticity-enhanced mechanism persists with additional HISs. In both cases ($N=3$ and $4$) the STAY-pathway fidelity reaches as high as 0.99997 mainly due to the increased adiabaticity ($\propto \mu$) for the dark state evolution.
Meanwhile, the optimized gate duration becomes even shorter as $N$ increases, consistent with the effective two-photon Rabi frequency analysis for TRANSFER pathway that is also $k$-dependent. This fact indicates that our protocol effectively overcomes the increasing multi-intermediate decay errors thereby the overall gate fidelity even achieves a small enhancement for a larger $N$.

{\it Evaluation with real Rubidium Atoms.-} In addition, we briefly discuss the feasibility of our protocol using Rb atoms. As described in Table~\ref{tab:k_factors}, we have identified a practical hyperfine-level configuration in Rb atoms that also satisfies the $k$ factor condition $k_0=k_1=k_r\equiv k$, with $k=0.577$. So we theoretically carry out 10 independent optimization runs for Rb atoms and present the results of a specific example in Table \ref{tab:multi_state} (last column). It is evident that the optimization procedure contains very slight randomness as the average fidelity $\bar{\mathcal{F}}_0$ 
converges close to 0.999 with fluctuations of only $\sim 6.2\times 10^{-6}$. Compared to the Cs case, we find the gate fidelity for Rb atoms is slightly lower because of the larger decay rate $\gamma_p$ along with smaller $k$ factors. Nevertheless, achieving enhanced adiabatic gates in Rb platforms remains possible, suggesting the broad applicability of our protocol to commonly-used Rydberg experimental platforms.

\bibliography{apssamp}

\end{document}